\documentclass[runningheads,a4paper]{llncs}

\usepackage{amssymb}
\usepackage{graphicx}
\usepackage{amsmath}
\usepackage{url}

\newcommand{\keywords}[1]{\par\addvspace\baselineskip
\noindent\keywordname\enspace\ignorespaces#1}

\begin{document}

\mainmatter  

\title{Machine learning based data mining for Milky Way filamentary structures reconstruction}

\titlerunning{Milky Way filament reconstruction with machine learning}

%
%
\author{Giuseppe Riccio\inst{1}%
\thanks{corresponding author, riccio@na.astro.it}%
\and Stefano Cavuoti\inst{1}\and Eugenio Schisano\inst{2}\and Massimo Brescia\inst{1}%
\and Amata Mercurio\inst{1}\and Davide Elia\inst{2}\and Milena Benedettini\inst{2}\and Stefano Pezzuto\inst{2}\and Sergio Molinari\inst{2}\and Anna Maria Di Giorgio\inst{2}}
\authorrunning{Riccio et al.}

\institute{INAF, Astronomical Observatory of Capodimonte, Via Moiariello 16, I-80131 Napoli, Italy\\
\and
INAF Institute of Space Astrophysics and Planetology, Via Fosso del Cavaliere 100, I-00133 Roma, Italy}

%
%

\toctitle{Lecture Notes in Computer Science}
\tocauthor{G. Riccio et al.}
\maketitle

\begin{abstract}
We present an innovative method called FilExSeC (Filaments Extraction, Selection and Classification), a data mining tool developed to investigate the possibility to refine and optimize the shape reconstruction of filamentary structures detected with a consolidated method based on the flux derivative analysis, through the column-density maps computed from Herschel infrared Galactic Plane Survey (Hi-GAL) observations of the Galactic plane. The present methodology is based on a feature extraction module followed by a machine learning model (Random Forest) dedicated to select features and to classify the pixels of the input images.
From tests on both simulations and real observations the method appears reliable and robust with respect to the variability of shape and distribution of filaments. In the cases of highly defined filament structures, the presented method is able to bridge the gaps among the detected fragments, thus improving their shape reconstruction. From a preliminary \textit{a posteriori} analysis of derived filament physical parameters, the method appears potentially able to add a sufficient contribution to complete and refine the filament reconstruction.
\end{abstract}

\section{Introduction}

The formation of star clusters is one of the most important events strictly related with the internal evolution of galaxies. Massive stars are responsible for the global ionization of the Interstellar Medium (ISM). About half of the mass in the ISM of a Galaxy is mainly derived by Molecular Clouds (MC) formed by stars. In such formation scenarios turbulent dynamics produce filamentary structures, where the total amount per unit area of suspended material measured along the length of a column (hereafter column density) agglomerates material from the Interstellar radiation field \cite{padoan2002}. As a natural consequence of the cooling process and gravitational instability, the filaments are fragmented into chains of turbulent clumps which may engage the process of star formation. Therefore, the knowledge about the morphology of such filamentary structures is a crucial information to understand the star formation process.

The traditional method, which represents our starting point in the design and application of the presented methodology, carries out filaments detection by thresholding over the image of the multi-directional 2nd derivatives of the signal to identify the spine, the area and the underlying background of the filaments, automatically identifying nodal points and filament branches \cite{schisano2014}. From the extracted regions of detected filaments it is possible to estimate several physical parameters of the filament such as the width, length, mean column density and mass per unit length.

In the present work we present a preliminary study of a data mining tool designed to improve the shape definition of filamentary structures extracted by the traditional method. This study is included into the EU FP7 project \textit{ViaLactea}, aimed at exploiting the combination of new-generation Infrared-Radio surveys of the Galactic Plane from space missions and ground-based facilities, by making an extensive use of 3D visual analysis and data mining for the data and science analysis, \cite{molinari2014}.

\section{The data mining approach}

\begin{figure}[!t]
\centering
\includegraphics[width=8cm]{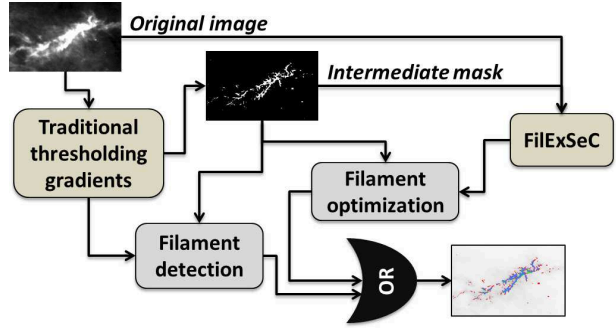}
\caption{Layout of the complete filamentary structure detection system.}
\label{fig:scheme}
\end{figure}

The presented data mining methodology, called FilExSeC (Filaments Extraction, Selection and Classification), is based on a feature extraction framework, able to map pixels onto a parameter space formed by a dynamical cross-correlation among pixels to produce a variety of textural, signal gradient and statistical features, followed by a Machine Learning (ML) supervised classification and backward feature selection, both based on the known Random Forest model \cite{breiman2001}. As shown in Fig.~\ref{fig:scheme}, the method is intended as an additional tool inserted into a more complex filament detection pipeline, whose role is to complement traditional consolidated techniques to optimize the overall performance.

FilExSeC is characterized by three main processing steps: (\textit{i}) encoding of pixel information into a data vector, formed by a set of derived features enclosing the information extracted from textural, flux gradient and statistical cross-correlation with neighbor pixels; (\textit{ii}) assigning the codified pixels to one of two filament/background classes; and (\textit{iii}) evaluating the importance of each feature in terms of its contribution to the classification task, with the purpose to detect and remove possible noisy features and to isolate the best subset of them capable to maintain the desired level of performance.

\subsection{The feature extraction}

The proposed method includes a Feature extraction algorithm, whose main target is to characterize pixels of the input image by means of Haar-like, Haralick and statistical features.
Haar-like features, \cite{lienhart2002}, are extracted by selecting windows $W_I$ of fixed dimension, where the pixel under analysis lies at the bottom-right vertex of the window. From these areas the integral image \textit{IntIm} is extracted, by setting the value of each point (x,y) as the sum of all grey levels of pixels belonging to the rectangle having pixel (x,y) and the upper-left pixel as vertices:

\begin{equation}
\text{IntIm}(x,y) = \sum \limits_{x' \le x,y' \le y} W_I (x',y')
\end{equation}

\begin{figure}[!t]
\centering
\includegraphics[width=10cm]{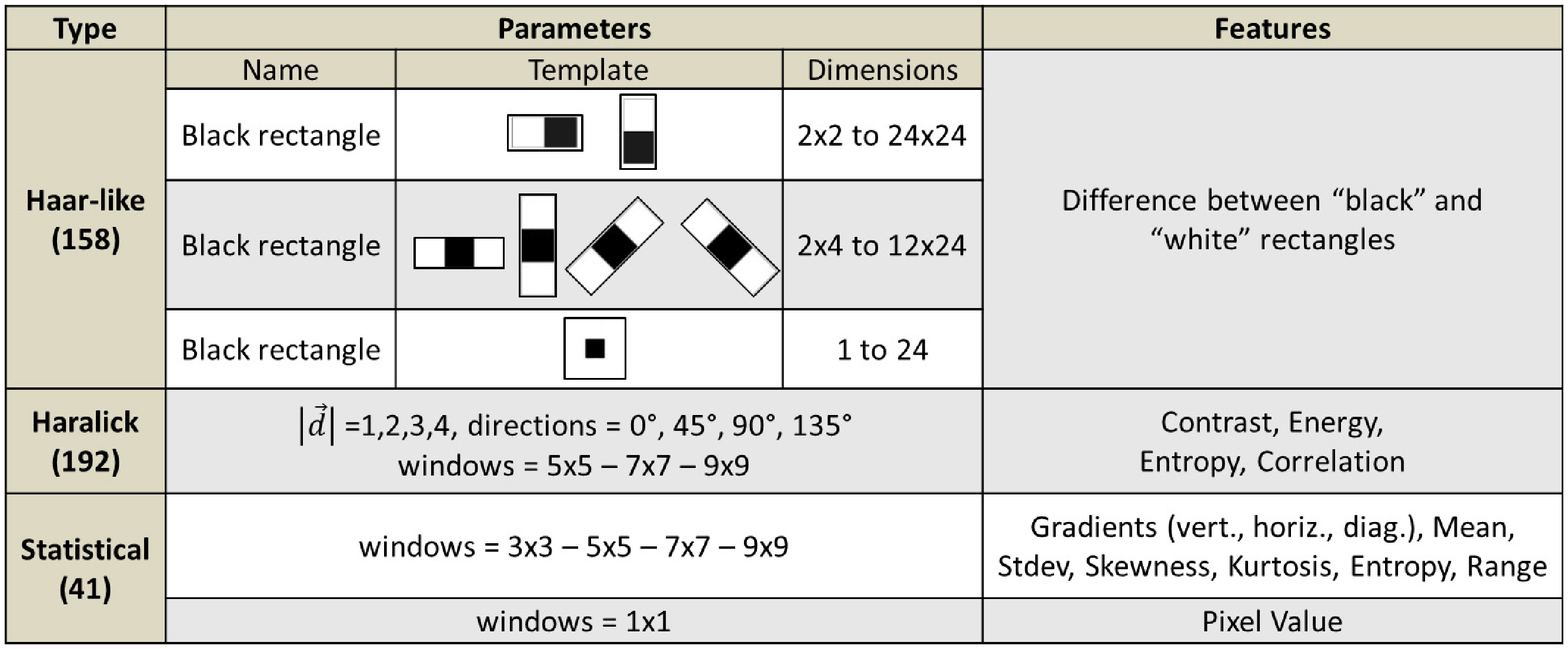}
\caption{Summary of the image parameter space produced by the feature extraction.}
\label{fig:features}
\end{figure}

Features are then calculated by comparing the content of the integral image among different regions, defined as \textit{black} and \textit{white} areas according to different templates specialized to search for different shapes in the image (Fig.~\ref{fig:features}):

\begin{equation}
f(x,y) = [\text{IntIm}(x,y)]_{black\: area} - [\text{IntIm}(x,y)]_{white\: area}
\end{equation}

Haralick features, \cite{haralick1979}, concern the textural analysis that, by investigating the distribution of grey levels of the image, returns information about contrast, homogeneity and regularity. The algorithm is based on the Gray Level Co-occurrence Matrix (GLCM), that takes into account the mutual position of pixels with similar grey levels. The $C_{i,j}$ element of the GLCM, for a fixed direction and distance $\overrightarrow{d}$, represents the probability  to have two pixels in the image at distance $\overrightarrow{d}$ and grey levels $Z_i$ and $Z_j$ respectively. Starting from the GLCM, it is possible to evaluate the most important Haralick features (Fig.~\ref{fig:features}), calculated within sub-regions with different dimensions and centered on the pixel under analysis, at different distances. Both dimensions and distances are parameters defined during setup.

The statistical features are calculated in sub-regions of varying sizes, centered on the pixel under analysis (Fig.~\ref{fig:features}).

\subsection{The classification and feature selection}

The feature extraction produces a list of features for all pixels, representing the input of the next steps.

The pixels belonging to filaments should have correlated values (or trends), although hidden by background noise. These values can be indeed used by a ML algorithm, hereinafter named as classifier, in order to learn how to discriminate the hidden correlation among filament pixels.

The classifier is based on the supervised ML paradigm. It has to be trained on a dataset (training set), where each pixel pattern has associated its known class label (for instance, filament or background). Then the trained model is tested on a blind dataset (test set), which again includes the known class for each pixel, in order to evaluate and validate the training and generalization performance. In the proposed method a Random Forest classifier has been used, \cite{breiman2001}.
After the validation test it is possible to proceed to the next step, i.e. the Feature Selection.
In general, the Feature Selection is a technique to reduce the initial data parameter space, by weighting the contribution (information entropy) carried by each feature to the learning capability of the classifier. By minimizing the input parameter space, it is hence possible to simplify the problem and at least to improve the execution efficiency of the model, without affecting its performance.
In principle, it is reasonable to guess that some pixel features could be revealed as redundant parameters, by sharing same quantity of information or in some cases by hiding/confusing a signal contribution.
Among the most known automated feature selection techniques we decided to use the Backward Elimination \cite{kohavi1997}, a technique starting from the full parameter space available, i.e. initially including all the given features. Then these are evaluated at each iteration in terms of contribution importance, dropped one at a time, starting from the least significant, and the KB is re-fitted until the efficiency of the model begin to get worse. Such technique has revealed to be well suited for image segmentation in other scientific contexts, \cite{tangaro2015}.
At the end of the feature selection phase, a residual subset of features with the higher weights are considered as the candidate parameter space. This subset is then used to definitely train and test the classifier. At the end of this long-time process the trained and refined classifier can be used on new real images.

\section{Experiments}

The presented method has been tested on the column density maps computed from Herschel \cite{pilbratt2010} observations of the Galactic plane obtained by the Hi-GAL project, which covers a wide strip of the inner Galactic Plane in five wavebands between $70\mu m$ and $500 \mu m$, (see \cite{elia2013} for further details). Here we report only one of the most illustrative examples due to problems of available space.

\begin{figure}[!th]
\centering
\includegraphics[width=12cm]{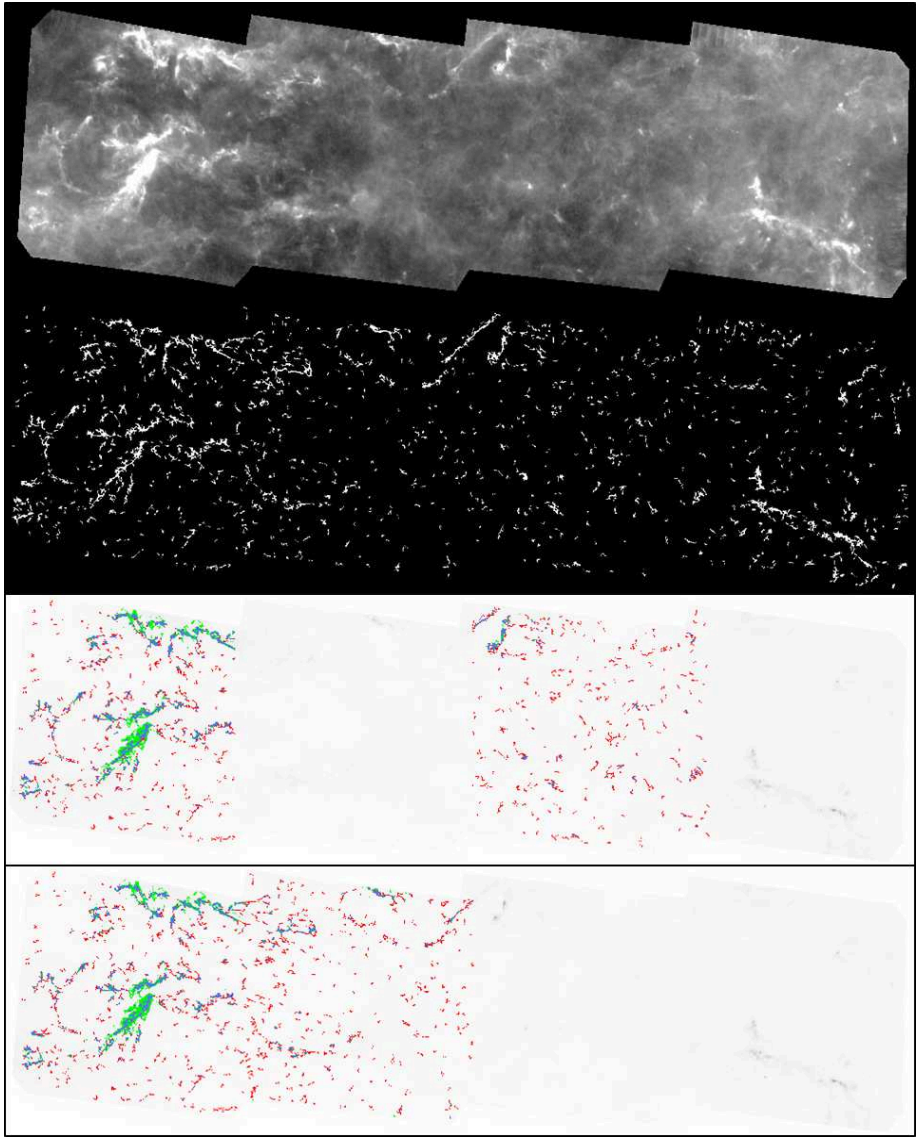}
\caption{The Hi-GAL region used in the reported example. It is a $2973\times1001$ pixels image of the column density map composed by $4$ tiles. Under the original observed image (upper side) the intermediate binary mask is shown. The other two sub-panels in reverse color show the output masks of FilExSeC in the two experiments, respectively, by using tiles $1$ and $3$ as blind test and by replacing tile $3$ with the tile $2$ in the test set. Here colored pixels are applied on the intermediate binary mask as classified by our method, representing, respectively, NFP (green) CFP (blue) and UFP (red) pixels.}
\label{fig:traintest}
\end{figure}

The data sample of the described test consists in a $2973\times1001$ pixels image of the column density map calculated from Hi-GAL maps in the Galactic longitude range from $l=217$ to $l=224^\circ$, \cite{elia2013}. In the reported example two different experiments are described. Their difference is related to the split of the four tiles to build up training and test sets.
As shown in Fig.~\ref{fig:traintest}, we assigned two different couples of tiles as training and test sets respectively, by always considering the forth tile as part of training set and the first one as part of the blind test set, while tiles $2$ and $3$ have been allocated alternately to training and test sets. The reasons of this strategy were on one hand to verify the reliability and robustness of our method with respect to the high variability in the shape and amount of filaments as well as of contaminating background distribution within the region pixels. On the other hand, by taking into account the basic prescription for the machine learning supervised paradigm, particular care has been payed in the selection of the training and test regions, trying to balance the distribution of different levels of signal and background between the two data subsets. It is in fact known that the generalization performance of a classifier strongly depend on the level of homogeneity and coherency between training and test samples. According to this strategy, tiles $1$ and $4$ are the two samples with the presence of the most relevant filament structures, while the other inner tiles are quite similar in terms of background distribution and low presence of filaments. Therefore, the two presented cases achieve the best balancing between training and test data.

In order to build up the knowledge base required by the supervised ML paradigm, the known target labels associated to the training and test patterns have been derived from an intermediate result of the traditional method, which assigns a binary label to each pixel, by distinguishing between filament or background pixel. Such intermediate binary masks are mainly composed by the central pixels of the filamentary regions. They are obtained by thresholding the eigenvalues map computed by diagonalizing the Hessian matrix of the intensity map of the region \cite{schisano2014}. These masks are still partially contaminated by non-filamentary structures (successively removed in the traditional approach by further filtering criteria), but the decision to use an intermediate product minimizes the bias introduced in our method by further filtering steps on the image samples.

The results of the pixel classification is represented in terms of the known confusion matrix, \cite{stehman1997}, in which the pixels are grouped in four categories:

\begin{itemize}
\item CFP (Confirmed Filament Pixel), filament pixels correctly recognized;
\item UFP (Undetected Filament Pixel), filament pixels wrongly classified;
\item NFP (New Filament Pixel), background pixels wrongly classified;
\item CBP (Confirmed Background Pixel), background pixels correctly recognized.
\end{itemize}

Some statistical indicators can be derived by the confusion matrix. For instance, we used the \textit{purity} (precision), i.e. the ratio between the number of correctly classified and total pixels classified as filament, the \textit{completeness} (recall), i.e. the ratio between the number of correctly classified and total pixels belonging to filaments and the \textit{DICE} ($F_1$-Score), which is a weighted average of purity and completeness. Such statistics have been mainly employed to verify the degree of reliability and robustness of the method with respect to the variability within the real images.

In the case of the two experiments of the presented example the results were, respectively, $\sim 74\%$ of purity, $\sim 52\%$ of completeness and $\sim 61\%$ of DICE in the experiment with test tiles $1$ and $3$ and $\sim 73\%$ of purity, $\sim 50\%$ of completeness and $\sim 59\%$ of DICE in the experiment with test tiles $1$ and $2$, for the filament pixels. While the statistics in the case of background pixel class were all enclosed around $98\%$ in both cases. These results appears quite similar in all the performed tests, with very small fluctuations, thus confirming what was expected, and by taking into account the extreme unbalance between the number of filament ($\sim 4\%$) and background ($\sim 96\%$) pixels. Furthermore, the FilExSeC method adds $\sim 16\%$ of new filament pixels (NFP) on average with respect to the intermediate image masks.

\section{Discussion and conclusions}

The aim of the innovative method is at improving the shape reconstruction of filamentary structures in IR images, in particular in the outer regions, where the ratio between the signal and the background is lower that in the central regions of the filament. So that the robustness of the method is evaluated for this particular science case, rather than in terms of an absolute detection performance to be compared with other techniques. The most important evaluation here consisted in the verification of an effective capability to refine the shape of filaments already detected by the traditional method, as well as to improve their reconstruction, trying to bridge the occurrences of fragments. We remind in fact, that our method starts from an intermediate result of the traditional method, for instance the intermediate binary masks obtained from the filament spine extraction, slightly enlarged through a method discussed in \cite{schisano2014}.

The proposed method revealed good reconstruction capabilities in presence of larger filament structures, mostly evident when the classifier is trained by the worst image regions (i.e. the ones with a higher level of background noise mixed to filament signal). This although in such conditions the reconstruction of very thin and short filament structures becomes less efficient, (Fig.~\ref{fig:traintest}). The global performances have been improved by optimizing Haar-like and statistical parameters as well as by introducing the pixel value as one of the features. Moreover, we have reached the best configuration of the Random Forest. Specific tests performed by varying the number of random trees have revealed the unchanged capability of classification with a relatively small set ($1000$ trees), thus reducing the complexity of the model.

\begin{figure}[!th]
\centering
\includegraphics[width=11cm]{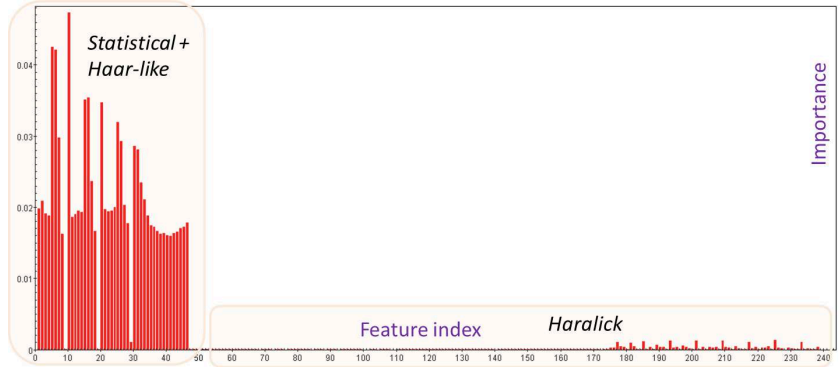}
\caption{Histogram of the feature importance as resulted from the feature selection with the Random Forest in the reported example. The highlighted regions show that within the $50$ most important features fall those of Haar-Like and statistical types. The Haralick features, located in the second region, resulted with a very low importance. This induced us to exclude them from the parameter space since their contribution is quite negligible, achieving also a benefit in terms of the reduction of computing complexity.}
\label{fig:frank}
\end{figure}

Furthermore, all the results of performed tests showed how the performance of the method worsen by eliminating the statistical features and that the Haralick features gave a very low contribution to the global efficiency. This was also confirmed by the feature importance evaluation (Fig.~\ref{fig:frank}), in which the textural features (in particular those of Haralick group) were always in the last positions. In this context, the importance of a feature is its relative rank (i.e. depth) used as a decision node of a tree to assess the relevance with respect to the predictability of the target variable. Features that are at the top of the tree induce a higher contribute to the final prediction decision for a larger fraction of the input patterns. The expected fraction of the patterns addressed by these features, can thus be used as an estimate of the relative importance of the features.

Therefore, the Random Forest feature importance analysis highlighted the irrelevance, in terms of information carriage, of Haralick type features, while confirming the predominance of statistical and Haar-Like types, independently from the peculiar aspects of both training and test images. The resulting minimized parameter space, obtained by removing Haralick features, had also a positive impact on the computing time of the workflow, since the $90\%$ of the processing time was due to Haralick parameters extraction.

Moreover, in case of training and test performed on a same image, like the presented example, it has been concluded that, by enhancing the variability of filament and background distributions in the training set, the method improves the purity and the completeness of filament classification. This can be motivated by the big variety of filament structures and by the extreme variability of background spread over the regions of an image. In fact, understanding where a filamentary structure merges into the surrounding background is the main critical point to be addressed, since it could help to determine its region extent\begin{center}
                                                                                                                                                                                                                                                                                                                                                                                                                                                                                                                                                                                                                                                  
                                                                                                                                                                                                                                                                                                                                                                                                                                                                                                                                                                                                                                                 \end{center}, as well as to realistically estimate the background. The latter is a fundamental condition to obtain a reliable determination of the filament properties, \cite{schisano2014}.

In order to evaluate and validate the results from the physical point of view, it is necessary to estimate the contribution of extended filamentary regions to the calculation of the filament physical parameters on the same regions, through a cross analysis of such physical quantities with the results obtained by the traditional technique. In particular, it is important to evaluate how the physics of filaments is strictly related to its mass and the contribution of the NFP pixels in the calculation of this quantity.
The variation of filament integrated column density is due to the contribution of the NFPs. In fact, it must be considered that the change of the distributions of filament/background mass contributions introduced by our method, causes an effect of a variation of the mass when NFPs are introduced and this is one aspect subject to a future further investigation.
Furthermore, as also visible in Fig.~\ref{fig:traintest}, in many cases our method is able to connect, by means of NFPs, filaments originally tagged as disjointed objects. In such cases, by considering two filaments as a unique structure, both total mass and mass per length unit change, inducing a variation in the physical parameters of the filamentary structure. However, an ongoing further analysis is required to verify the correctness of the reconstruction of interconnections between different filaments, and to better quantify the contribution of FilExSeC to the overall knowledge of the physics of the filaments.

\subsubsection*{Acknowledgments} \noindent This work was financed by the 7th European Framework Programme for Research Grant FP7-SPACE-2013-1, \textit{ViaLactea - The Milky Way as a Star Formation Engine}.

\end{document}